# W3-Scrape - A Windows based Reconnaissance Tool for Web Application Fingerprinting

Karthik R[a], Raghavendra Karthik [a], Pramod S [a] * and Sowmya Kamath[a]

[a] *Department of Information Technology, National Institute of Technology Karnataka,*
*Srinivas Nagar, Surathkal, Mangalore 575025, INDIA*

**Abstract**

Web Application finger printing is a quintessential part of the Information Gathering phase of (ethical) hacking. It allows narrowing down the specifics instead of looking for all clues. Also an application that has been correctly recognized can help in quickly analyzing known weaknesses and then moving ahead with remaining aspects. This step is also essential to allow a pen tester to customize its payload or exploitation techniques based on the identification so to increase the chances of successful intrusion. This paper presents a new tool "W3-Scrape" for the relatively nascent field of Web Application finger printing that helps automate web application fingerprinting when performed in the current scenarios.

*Keywords:* Web Application Fingerprinting; ethical hacking; exploitation technique; Security;

## 1. Introduction

Finger printing, in the simplest sense, is a method used to identify objects. The term has been used to identify TCP/IP Stack Implementation and was known as TCP/IP finger printing. This approach has been extended to identify web application specifics. Web Application finger printing is performed to identify the application and software stacks running on the HTTP Server. Currently, Web application finger printing is at its embryonic stage, however interest in this field is rising steadily and large number of automated solutions are emerging in the market.

HTTP Fingerprinting [1] is a technique that helps determine details like the web server software hosting the website, its version and other deployment details of the web server. HTTP fingerprinting allows network administrators to profile the web servers in their environment and monitor patches. It also allows a pen-tester/security auditor to narrow down the list of attacks that the server must be subjected to, to expose vulnerabilities. There are several vendors in the market today like Microsoft, Apache, Netscape and others; their products differ in the ways in which they implement the HTTP protocol. Unfortunately, this is also the reason why HTTP Fingerprinting becomes possible and may be used for nefarious purposes. [1]

Every company with a web presence opens the TCP port 80 on their firewalls to the Internet for web-based applications. Web servers can reveal very juicy information that attackers can feast up on. This helps them to refine the attack the plan. The patch information, the application running on the web server can be easily got from this. Blindly attacking may lead to detection by IDS as the security vulnerabilities are vendor specific. Knowing the above gathered information can greatly improve the efficiency of the attack. Proper usage of exploits increases the chance of successful exploitation. [2]

In the present day, a lot of web application vulnerabilities are researched and published for various CMS's. Topping the list is the SQL injection and XSS (Cross site scripting). SQL injection is of two types: Blind SQL injection and Classical SQL injection. The difference between the two is the presence of *information_schema* in the later versions. This can give administrative access to the web application if properly executed. Hence, addressing the SQL vulnerability of the target web application is an important aspect of our tool and has been added as an extension. There are two types of XSS attacks - Persistent and reflected. The former is more dangerous than the latter as it can modify server side data. But the latter can be used in one of the more popular Social engineering attacks - Phishing. The URL redirection script embedded on the cross site script can make the victim believe that the phished page is indeed arising from a legal vendor and hence may be lured to click on the suspicious link. To rule out most of the suspicious links, we have also added the Phishing detection tool for this framework.

Hence, web application fingerprinting is the most important stage in the penetration testing of Web applications. Our tool comes in handy in the Reconnaissance phase of the Hacker cycle. Fingerprinting the target allows us to narrow down the specifics instead of looking for all possible permutations and combinations of data. For any successful attack to happen, it is very important that the target vulnerabilities are known and also the feature corresponding to that particular vulnerability is found. Our tool helps in the vulnerability assessment phase as it provides close proximity information about the who's who of the target. This in turn helps in improving the defense against any such future attacks. The tool also has a few other extensions like preliminary SQL injection tester and a Phishing detection extension module as a part of this framework, which, to our knowledge, none of the tools mentioned earlier offer.

This paper is organized as follows – Section 2 presents a discussion in this area based on existing literature; Section 3 provides details about our experimental setup, methodology and implementation specifics highlighting the working of our tool and the results. Section 4 provides a comparison of W3-Scrape with existing tools. Section 5 presents identified future work followed by conclusion in Section 6 and a list of references.

## 2. Literature Survey

Developing applications that run in the distributed operating environment of today's Internet requires an efficient, easy-to-use method for retrieving data from resources of all types. Pluggable protocols help in developing applications that use a single interface to retrieve data from multiple Internet protocols. [2] Methods that are already in use for this purpose include HTML Data Inspection, File and Folder Presence (HTTP response codes) and Checksum Based identification. [3] Various tools available currently that are based on these assumptions are WhatWeb, Wapplyzer, BlindElephant, Plecost and W3af Wordpress finger printer. Below is brief discussion on the technical aspects of these tools.

*2.1. Tools on the market*

WhatWeb is a Ruby based application allowing a pluggable architecture with virtually any application detection. It performs tasks like Google dork check, Regex pattern matching, File existence checker, File Content checker based on file name and MD5 based matching. All this effectively allows WhatWeb to report the application's status more accurately. The fact that it is pluggable in nature allows it to be customized for any application encountered.

Wapplyzer is a Firefox/Chrome Plugin, and works on only regular expression matching and doesn't need anything other than the page to be loaded on the browser. It works completely at the browser level and gives results in the form of icons.

BlindElephant is a new entrant in the market and works on the principle of static file checksum based version difference. This allows this software to work for both open-source software and closed source platforms, the only condition being - the person running BlindElephant needs to have access to source code to map all static file fingerprinting. The technique basically works by - create checksum local file and store in DB, download static file from remote server, create checksum and ompare with checksum stored in DB and identify it.

Plecost works on a simple principle of finding right files. It derives the version of Wordpress from readme.html of the website. Basically it tries to fetch the readme.txt for each plugin and then based on that deduces the version of appliance installed on this server. Since Wordpress makes it mandatory for every plugin author to have a correctly formed readme.txt file so that chances of finding these files are very high.

W3af is attracting quite a lot of attention nowadays. This plugin takes a retro approach, looking for exact file names /paths and moving on to look for specific data inside the file and if existing, then deduce that the application is Wordpress. This highlight is to stress on the fact about paths and flaws.

*2.2. Motivation*

The inherent flaws in the design of current automation tools are that these tools work on the basis of assumptions. For example they assume that - Filename X means Z plugin; or Folder Q in site means software is used irrespective of the actual usage of the product or not and they proclaim the website to be dynamic however it remains largely static. They claim to be dynamic however remain static at large parts. (dynamic only in replacing the front domain name). The results are based on presence of Static paths and dependency on static filenames. Even the checksum based approach suffers with these limitations, which turns out to be a fundamental flow in this approach. [7] For a security analyst, it's a key requirement to perform phase 1 of penetration testing in a great detail. Similarly, here, foot printing of web applications is a necessary starting point in any of the web application security analysis.

**3. Methodology and Implementation**

W3-Scrape is a reconnaissance based reporting tool that focuses on generating reports for web application fingerprinting - identifying the server and corresponding operating systems along with the Content management system running on a particular web application. The following section describes the functionality of each module and the implementation specifics. We considered the requirements of a Web application tester and tried to integrate the functionalities in one single composite tool.

W3-Scrape has the following components - HTML Source view, Harvesting IP address, Rendering the HTML Source, Website Scraping, CMS identification, SQL vulnerability identification, Phishing Detection and Full Report Generation. The System.Net namespace provides HttpWebRequest and HttpWebResponse classes. Sytem.IO namespace provides classes to send request and receive response in streams. This is the base class, which provides methods to request data, parse any redirection URL, receive response and convert response (since it's coming in streams) into meaningful data.

*3.1. HTML Source view*

This module grabs the HTML source code of the target URL and is used for manual analysis of the HTML Source code. The feature working is similar to "View page source" in any browser. It uses HttpWebResponse class to capture the page source which provides an HTTP-specific implementation of the WebRequest class by creating an object that contains all the relevant information required to generate a proxy used to communicate with a remote object and return a response from an Internet resource. Fig. 1(a) shows the tool generating the HTML source view.

*3.2. Harvesting IP address*

A harvester is a computer program that surfs the internet looking for email addresses. Harvesting email addresses from the Internet is the primary way spammers build their lists. Harvesters must connect to the Internet through an IP address. We use the IPHostEntry Class to find any IP address harvesting vulnerabilities. The class IPHostEntry provides a container class for Internet host address information, gets or sets a list of IP addresses that are associated with a host and gets or sets the DNS name of the host as shown in Fig. 1(b)

*3.3. Rendering HTML source*

This uses the Web Browser Class for its working. It loads the document at the location indicated by the specified URI into the Web Browser control, replacing the previous document and reloads the document currently displayed in the Web Browser control by checking the server for an updated version. Fig. 2(a) shows the rendered HTML page. This doesn't serve any important purpose, but makes the system user friendly to browse through the target web site well within the framework instead of viewing it in a separate browser. This enhances the UI and also the ease of use of this framework with respect to the user.

*3.4. Web Scraping*

Web Scraping feature utilizes the HttpWebResponse Class which provides an HTTP-specific implementation of the WebResponse class and captures HTTP header information and fingerprints the following: Content- Length,

Content- Type and the server on which the host is launched. The information that we get from this module becomes the most important aspect of the attack. This provides the attacker/tester the necessary information from which he can narrow down his search criteria in the Vulnerability assessment phase of the cycle. He can directly probe in to the vulnerabilities of the particular Operating system run by the system rather than scrounging for unnecessary information.

For example, if the tester finds the target application as Cent OS, then it saves him time from finding all the vulnerabilities one-by-one for all other Operating systems that might probably support the CMS of the target. Another important use is to know the server on which this particular CMS is hosted, thus he can replicate the environment in his local area, to use it for simulating the target environment and observe the application behavior locally. Thus, this particular module turns out to be very handy. Fig. 2(b) shows the Web scrape module.

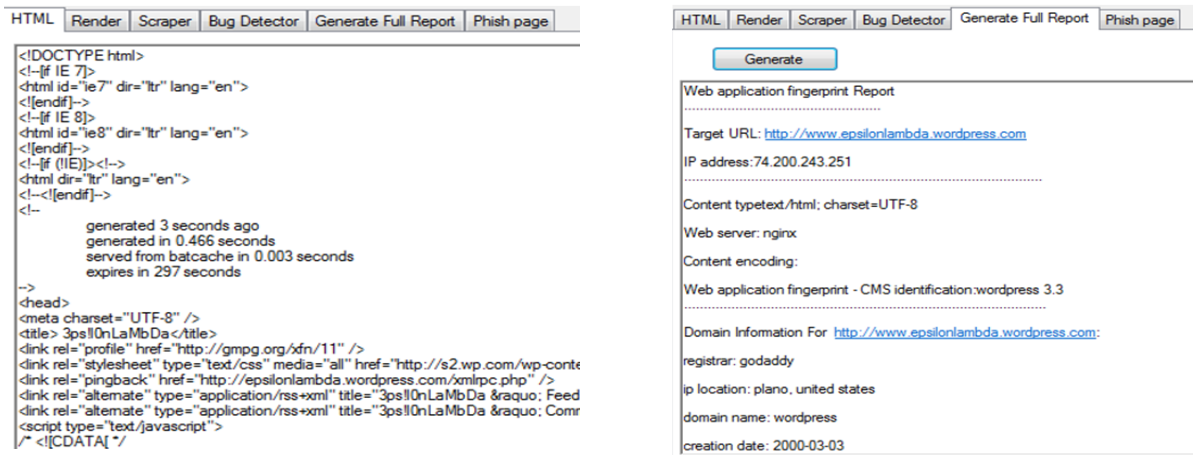

Fig.1. (a) HTML Source view generated by W3-Scrape   (b) IP Address and other details harvested from the CMS

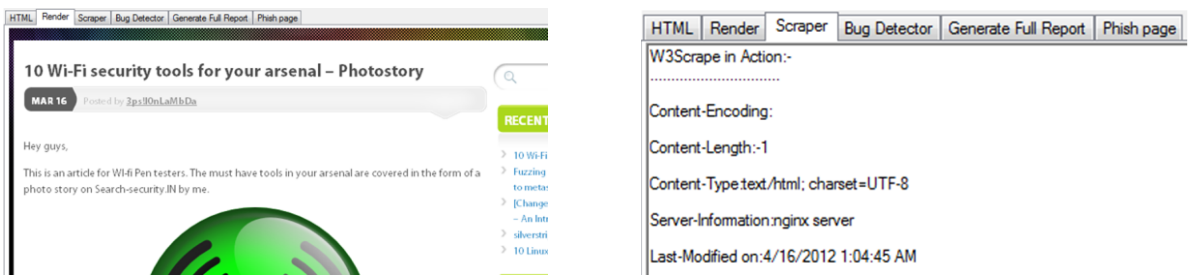

Fig. 2. (a) Rendering the webpage within the W3-Scrape framework   (b) The Web Scrape Module

*3.5. CMS identification*

In our approach of CMS identification, we use the HTML data inspection method as well as URL inspection methods. Here, regular expressions and pattern matching plays a crucial role in the success of the prediction. There may be some non-CMS based websites where a system based error is thrown by the application. But, the above said methodology works on majority of the sites tested by us manually. Figure 1(b) above also shows W3-Scrape retrieving the name of the CMS and related details.

*3.6. SQL injection vulnerability detection*

Our application detects the SQL injection bug, and also suggests the sanitization of the attacking point. The bug detector module detects security bugs like SQL injection vulnerability in the URL, and also the Local File injection Vulnerability of any site. We use regular expressions to match certain criteria of the test and verify the vulnerability of the target. [5]

Fig. 3(a) shows a SQL vulnerable site detected by our bug detector. We have tested around 100 sites for this purpose, and W3-Scrape has detected every site with a SQL error without generating any false positives. The reason may be that, any site with SQL errors will throw up an error message on appending a 'quote' to the URL. Thus, our system catches the server throw to read through the error and then deliver the result.

Another bug detection method incorporated here is the Local File Injection Vulnerability. A LFI is a method for servers/scripts to include local files on run-time, in order to make complex systems of procedure calls. LFI vulnerabilities are most often found in URLs of the web pages, mainly because developers tend to use GET requests when including pages. When we try to tamper with the URL, we can change the URL equations and determine if there is a Local File Injection Vulnerability, when an attack string is appended to the URL. LFI has really devastating consequences, few of them being Poison of Null byte, Log Poisoning, malicious image upload etc. The ability to detect such bugs will give administrators a chance to secure the site before it is attacked.

*3.7. Phishing Detection*

W3-Scrape includes a phishing detection module that is not available in any of the tools discusses in Section II. Determining if a page is phished can be done based on different factors such as usage of IP based URLs, too many dots in the URL, age to linked-to domain names etc. Fig. 3(b) shows a screenshot of this feature.

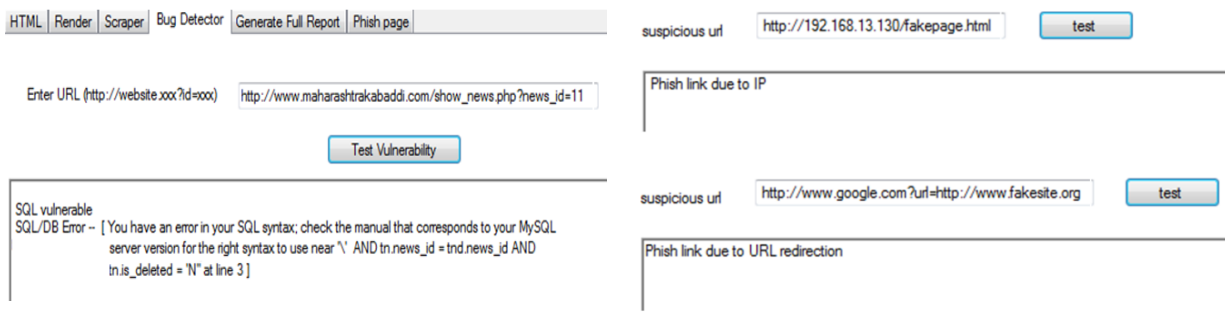

Fig. 3. (a)Bug Detector Module    (b) Phishing Detection.

*3.8. Full Report Generation*

The Generate full report module prints the entire fingerprinting report in the panel. The report consists of the target URL, IP address, content type, and Web server type, Content Management system followed by the domain specific information such as registrar information, IP location, domain name, creation date and Expiry date. This report will be followed by the Vulnerability report. Fig. 4 shows the full report generated by the tool.

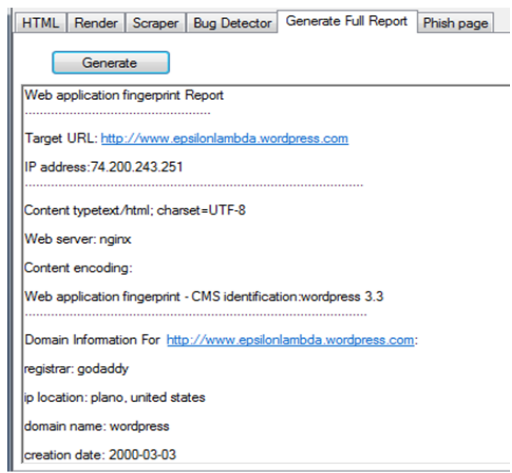

Fig. 4. Full Report Generation in W3-Scrape

## 4. Comparison with other tools currently available

The W3-Scrape framework uses Windows APIs for all the reporting and analysis. Thus, the primary advantage is that, it caters to the needs of people not familiar with Linux machines. Secondly, this project has a very simple to use and easy to understand GUI, which all the other tools in the market lack. An enhanced GUI not only enhances the user experience but also speeds up the learning of the tool. Next, the reporting functionality provides a variety of information in one single framework. No other tool in the market provides all this information in one single place as W3-Scrape. Table 1 presents a comparison of W3-Scrape and the other tools.

Table 1: Comparison of W3-Scrape with other tools available on the market.

| Tool | GUI | Bug Detection (LFI and SQLi) | Phishing Detection | Reporting Abilities | Who-is Information |
|---|---|---|---|---|---|
| BlindElephant | No | No | No | Yes – Plaintext | No |
| Wapplyzer | Browser Addon | No | No | Yes – Icons | No |
| Plecost | No | No | No | Yes – Plaintext | No |
| Whatweb Wordpress Fingerprinter | No | No | No | Yes – Plaintext & XML | No |
| W3af | No | No | No | Yes – Plaintext & XML | No |
| **W3-Scrape** | **Yes** | **Yes** | **Yes** | **Yes – Plaintext** | **Yes** |

## 6. Conclusion

W3-Scrape can be used by both white hats as well as black hats based on the motive of their actions. There are modules that fetch readily available public data from the Internet using various online resources. This makes the tool handy for a penetration tester to narrow down the possibilities of the target application. Finger printing in its simplest senses is a method used to identify objects. Same term has been used to identify TCP/IP Stack Implementation and was known as TCP/IP finger printing. Similar usage has been extended lately to identify web applications installed on the Http Server. With all additional modules integrated along with reconnaissance modules, this tool has really good potential to be a well rounded framework in the future, however helping attackers as well as "ethical" hackers equally.